\documentclass[twocolumn,pra,superscriptaddress]{revtex4-1}

\usepackage{bm}%
\usepackage{float}
\expandafter\ifx\csname package@font\endcsname\relax\else
 \expandafter\expandafter
 \expandafter\usepackage
 \expandafter\expandafter
 \expandafter{\csname package@font\endcsname}%
\fi

\usepackage{array}
\usepackage{amsmath,amssymb}
\usepackage{MnSymbol}
\usepackage{tikz-feynman,contour} 
\usetikzlibrary{shapes,arrows,positioning,automata,backgrounds,calc,er,patterns}
\usepackage{feynmf}
\tikzfeynmanset{compat=1.0.0} 
\usepackage{graphicx}
\usepackage{mathrsfs}
\usepackage{bm}
\usepackage{mathtools}
\usepackage{epstopdf}
\usepackage{hyperref}

\usepackage{amsmath}
\usepackage{amsfonts}
\usepackage{mathtools}
\usepackage{graphicx}
\usepackage{fixme}
\usepackage{color}

\usepackage{wasysym}

\begin{document} 
\title{Strong photon interactions from weakly interacting particles}
\author{A. Camacho-Guardian}
\affiliation{Departamento de F\'isica Qu\'imica, Instituto de F\'isica, Universidad Nacional Aut\'onoma de M\'exico, Apartado Postal 20-364, Ciudad de M\'exico C.P. 01000, Mexico}
\author{M. Bastarrachea-Magnani}
\affiliation{Departamento de F\'isica, Universidad Aut\'onoma Metropolitana-Iztapalapa, San Rafael Atlixco 186, C.P. 09340 CDMX, Mexico.}
\affiliation{Center for Complex Quantum Systems, Department of Physics and Astronomy, Aarhus University, Ny Munkegade 120, DK-8000 Aarhus C, Denmark.}
\author{T. Pohl}
\affiliation{Center for Complex Quantum Systems, Department of Physics and Astronomy, Aarhus University, Ny Munkegade 120, DK-8000 Aarhus C, Denmark.}
\author{G. M. Bruun} 
\affiliation{Center for Complex Quantum Systems, Department of Physics and Astronomy, Aarhus University, Ny Munkegade 120, DK-8000 Aarhus C, Denmark.}
\affiliation{Shenzhen Institute for Quantum Science and Engineering and Department of Physics, Southern University of Science and Technology, Shenzhen 518055, China. }

\begin{abstract}
 The hybridization of light and matter excitations in the form of polaritons has enabled major advances in understanding and controlling optical nonlinearities. Entering the quantum regime of strong interactions between individual photons  has however remained challenging since the strength of achievable polariton interactions is typically limited by the available interactions in the material.
Here, we investigate collisions between dark-state polaritons in three-level systems and discover a resonant process that yields effective interactions, which are much larger than the underlying interaction between their matter constituents. We systematically investigate the underlying mechanism and demonstrate a substantial enhancement of polariton interactions by several orders of magnitude. This suggests a promising approach to quantum nonlinear optics in a range of physical settings, from atomic gases to excitons in semiconductors and two-dimensional bilayer materials.
\end{abstract}
\date{\today}
\maketitle


When light couples to a material it can form polaritons that are composed of photons and collective matter excitations. Owing to this hybrid character, polaritons inherit the ability to interact from their matter-constituents. This gives rise to a host of exciting nonlinear optical phenomena \cite{Carusotto2013}, from the realization of optically controlled memories \cite{Cerna2013}, and the generation of squeezed light \cite{Boulier2014}, to the observation of polariton condensation \cite{Kasprzak2006} and superfluidity \cite{Amo2009}, as well as dissipative phase transitions \cite{Fink2018}. However, the interactions available in most materials are generally too weak to generate nonlinearities that are strong enough to act at the ultimate level of single photons. In light of the great scientific and technological significance of such effective photon interactions \cite{Chang2014}, substantial ongoing efforts are being directed towards finding materials with large particle interactions that can be exploited for optical applications \cite{Wouters2007,Carusotto2010,Takemura2014,Peyronel2012,MURRAY2016,Rosenberg2018,Thompson2017,Togan2018,Walther2018,Emmanuele2020,Tan2020,Bastarrachea-Magnani2021,Zhang2021,Camacho2021}.

\begin{figure}[t!]
\begin{center}
\includegraphics[width=\columnwidth]{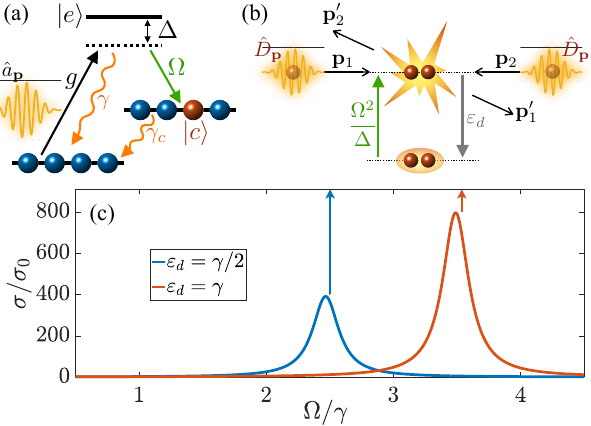}
\end{center}
\caption{Optical nonlinearities in an interacting three-level medium. (a) Underlying three-level coupling scheme. The incident probe light, described by operators $\hat{a}_{\bf p}$, interacts with the medium via a two-photon transition and creates $|c\rangle$-excitations. On two-photon resonance, electromagnetically induced transparency enables  low-loss propagation of the probe light in the form of dark-state polaritons, described by $\hat{D}_{\bf p}$. (b) The $|c\rangle$-state excitations interact via a weak short-range interaction with an effective polariton interactions cross-section $\sigma$. The existence of a dimer state at a binding energy $\varepsilon_d$ can lead to a resonance where the polariton interaction exceeds that of the bare excitations. (c) Polariton scattering cross-section in units of this bare value, $\sigma_0$ for $\Delta=-25\gamma$, $\gamma_c=0.015\gamma$, $g=1000\gamma$ and two indicated dimer energies. The vertical arrows mark resonance position as predicted by Eq.~\ref{ResCon}.} 
\label{Fig1}
\end{figure}

Here, we describe a general mechanism that overcomes this limitation and makes it possible to generate strong effective polariton interactions in a weakly interacting optical medium. This surprising effect occurs in three-level systems under resonant  two-photon coupling between incident photons and long-lived matter excitations. It emerges from an optical resonance with a deeply bound dimer state formed by two of such excitations as illustrated in Figs.\ref{Fig1}(a) and (b), and differs from a collisional polaritonic Feshbach resonance where the collisional energy of the colliding pair is tuned close to the binding energy of the dimer. Contrary to this, we show that 3-level dark-state polaritons can be created on resonance around zero collision energies, while the bound state resides at a large negative energy and is lifted into resonance by an external control field. 
Upon optimizing the effect for short-range interactions, we find a polariton scattering length that scales as $\sim 1/a$ with the scattering length, $a$, of the colliding excited particles. We will show that this implies a vast enhancement of polariton interactions by several orders of magnitude when the interaction between the particles is weak. 

Figure \ref{Fig1} illustrates the basic setup in which an optical medium is driven by weak probe light described by the photon creation operators $\hat{a}_{\bf p}^\dagger$ with momentum ${\bf p}$. 
The absorption of photons in the optical medium generates excitations to the state  $|e\rangle$ 
 as expressed by another set of bosonic creation operators $\hat{e}_{\bf p}^\dagger$. 
 A  stronger classical control field with a wave vector  ${\bf k}_c$ and frequency $\omega_c=c k_c$ couples these
  excitations to another  state $|c\rangle$  with associated creation operators $\hat{c}_{\bf p}^\dagger$. The Hamiltonian is
  
 \begin{align}
\hat H=&\sum_{\mathbf p}\left(E^{e}_{\mathbf p}\hat e^\dagger_{\mathbf p}\hat e_{\mathbf p}
+E^{c}_{\mathbf p}\hat c^\dagger_{\mathbf p}\hat c_{\mathbf p}+E^{a}_{\mathbf p}{\hat a}_{\mathbf p}^\dagger\hat{a}_{\mathbf p}\right)+\nonumber\\
&
+\sum_{{\mathbf p}}( g\hat e^\dagger_{\mathbf p} {\hat a}_{\mathbf p}+\Omega  \hat c^\dagger_{\mathbf p-\mathbf k_{\text{c}}}\hat e_{\mathbf p}+\text{h.c.}),
\label{Hamiltonian}
\end{align} 
concrete physical implementations will be discussed below. Here  $E^{e}_{\mathbf p}=\epsilon_{\mathbf p}+\epsilon_e-i\gamma$ and $E^{c}_{\mathbf p}=\epsilon_{\mathbf p}+\epsilon_c+\omega_c-i\gamma_c$
are the excitation energies of the $|e\rangle$ and $|c\rangle$ states, $\epsilon_{(e/c)}$ are the bare energies of the  $|e\rangle$ and $|c\rangle$ states. The kinetic energy of the  excitations is given $\epsilon_{\mathbf p}=p^2/2m$, where $m$ denotes their mass and we have used units for which $\hbar=1$. We include decay rates $\gamma$ and $\gamma_c$ of the 
 $|e\rangle$ and metastable $|c\rangle$ states with  with $\gamma_c\ll \gamma$. 
 The probe photons propagate with the linear dispersion $E^{a}_{\mathbf p}=c p$ at the speed of light $c$.
 While we for concreteness consider light propagating in a three-dimensional continuous atomic gas,  the formalism can be straightforwardly be adopted to two-dimensional lattice geometries relevant for layered semiconductors in microcavities. The single-photon coupling strength of the probe light is given by $g$ and $\Omega$ is the Rabi frequency for the control-field. 
 Note that $g\propto \sqrt n$ where $n$ is the 
 density of the optical medium formed by the ground state atoms. 
  In the absence of the control field coupling, $\Omega=0$, we have a simple two-level medium that absorbs light with a resonant absorption length $\ell=\gamma c/g^2$. A finite control field coupling, however, can lead to electromagnetically induced transparency (EIT)  such that the probe light can propagate without 
 losses~\cite{Fleischhauer2005,SM}.

This is readily seen from the eigenstates of Eq.~\eqref{Hamiltonian}  describing polaritons. An important case arises for photon frequencies $c p_r =E^{c}_{{\bf p}_r-{\bf k}_{c}}$, which facilitate the resonant two-photon generation of $|c\rangle$-excitations with an intermediate single-photon detuning $\Delta=E^{e}_{{\bf p}_r}-c p_r$. Around such resonant momenta ${\bf p}_r$, one finds a simple solution
\begin{align}
\hat D_{\mathbf p}\approx-\cos\theta {\hat a}_{\mathbf p}+\sin\theta \hat{c}_{\mathbf p-\mathbf k_{\text{c}}}
\label{DS}
\end{align}
 that corresponds to a dark-state polariton, in which photons hybridize with $|c\rangle$-excitations with a mixing angle $\tan\theta=g/\Omega$ and corresponding Hopfield factors $\sin\theta$ and $\cos\theta$. As the dark-state polariton does not contain the decaying $|e\rangle$-state, Eq.~\eqref{DS}  implies a virtually lossless propagation, due to EIT. Moreover, since the collective photon coupling $g$ is typically much larger than the control-field Rabi frequency $\Omega$, the polariton  contains a vanishingly small photon component, and, hence, propagates with a low group velocity $v_g=c\cos^2\theta \ll c$ \cite{Fleischhauer2002}. Correspondingly, the polariton is almost entirely composed of $|c\rangle$-excitations with an associated quasiparticle residue of $Z_c=\sin^2\theta\approx 1$. One may, thus, expect that this situation maximizes achievable optical nonlinearities, whereby the polaritons simply acquire the interactions between the $|c\rangle$-states~\cite{Fleischhauer2008}. 

\begin{figure}[t!]
\includegraphics[width=1\columnwidth]{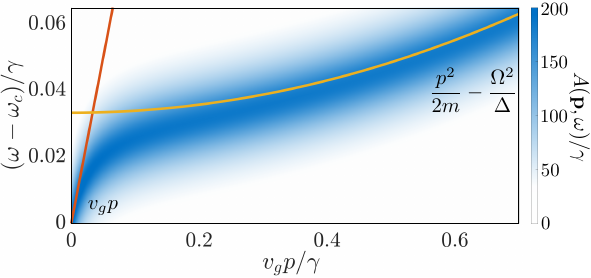}
\caption{The spectral function of the $|c\rangle$ state for $\Omega=\gamma$, $g/\Omega=25000$, $\Delta=-30\gamma$, $\gamma_c=\gamma/100$. The 
dark-state polariton with a linear dispersion, $\omega-\omega_c \sim v_g p$, turns into a dressed $|c\rangle$-state excitation with the quadratic dispersion, $\omega-\omega_c \sim p^2/2m-\Omega^2/\Delta$, as the photon decouples with increasing momentum $p$.  
}
\label{Fig2}
\end{figure}

Surprisingly, this simple perturbative picture, which works well for two-level systems, fails to capture the general physics of optical three-level media.
As we shall see below, the optical nonlinearity of such three-level settings can exceed the available interactions in the material by orders of magnitude. To this end, we use a non-perturbative description, starting from the Green's function 
\begin{equation}\label{Gc}
 G_{c}^{-1}({\mathbf p},\omega)=\omega-E^{c}_{\mathbf p}
- \frac{\Omega^2}{\omega- E^{e}_{{\mathbf p}+\mathbf k_{\text{c}}}-\frac{g^2}{\omega-E^{a}_{{\bf p}+{\bf k}_{\text{c}}}}},
\end{equation}
of the  dressed $|c\rangle$-state excitations. Equation \eqref{Gc} follows exactly from Eq.~\eqref{Hamiltonian}~\cite{Camacho2020}, and its spectral function 
$A_{c}(\mathbf p,\omega)=-2\text{Im}{G}_{c}(\mathbf p,\omega)$ sharply peaks around the dispersion of the polariton solutions. 
The spectral function shown in Fig.~\ref{Fig2} clearly reveals the familiar linear slow-light group dispersion $\omega-\omega_c\approx v_g p$ for sufficiently low momenta.
Away from the EIT condition, the photons, however, decouple due to their 
 steep dispersion and the energy of the polariton becomes $q^2/2m-\Omega^2/\Delta$, if $|\Delta|\gg \gamma$. This corresponds to the limit of a bare $|c\rangle$-excitations with kinetic energy, $p^2/2m$, while the weak optical dressing by the control field generates a light shift $-\Omega^2/\Delta$ at high momenta~\cite{Nielsen2020}. As we will show,  it is this distinctly changing character of the polariton dispersion that enables the enhancement of photon-photon interaction.

Upon scaling lengths by $\ell$ and frequencies by $\gamma$ we are left with 5 dimensionless parameters, $g/\gamma$, $\Omega/\gamma$, $\Delta/\gamma$, $\gamma_c/\gamma$, and $c p_r/\gamma$, that determine the polariton propagation in the system. In addition, this scaling of Eq.(\ref{Gc}) gives two constants
$c k_{\text{c}}/\gamma$ and $\gamma/mc^2$, which we for concreteness fix  for an atomic gas of 
rubidium atoms \cite{Rb87}.

Let us now proceed by exploring the effects of interactions between the photon-generated excitations. Since the dark-state polariton has only marginal contributions from the $|e\rangle$-states, we can focus on the interactions between the $|c\rangle$-states.
Interactions involving the ground state atoms are assumed to be weak and can be included via mean-field shifts if needed.
Consider the scattering of two dark-state polaritons from momenta ${\bf p}_1$ and ${\bf p}_2$ to ${\bf p}_1^\prime$ and ${\bf p}_2^\prime$, under EIT conditions, $p_i=p_i'=p_r$, see Fig.~\ref{Fig1}(b). The collision  is solely driven by the interaction between the 
$|c\rangle$-state components of the polaritons, which are given by Eq.~\eqref{DS} and, thus, move with respective momenta ${\bf p}_i-{\bf k}_c$. For a given center of mass momentum ${\bf p}_1+{\bf p}_2$ and total energy $2cp_r$ of the two colliding polaritons, the relevant scattering matrix is therefore given by 
${\mathcal T}=\sin^4\theta {\mathcal T}_{cc}(\mathbf p_1+\mathbf p_2-2{\mathbf k}_c,2cp_r)$, where ${\mathcal T}_{cc}$ is the scattering matrix 
for the dressed $|c\rangle$-state excitations, whose propagator is given by  Eq.~\eqref{Gc}. For a short range interaction 
with a negligible momentum dependence at the relevant energies, the Lippmann-Schwinger equation for the two-polariton scattering problem has the solution~\cite{Fetter1971}
\begin{align}
\label{Tm}
\mathcal T_{cc}^{-1}(\mathbf Q,\omega)=\mathcal T_0^{-1}-\int\!\frac{{\rm d}^3q}{(2\pi)^3}\Pi_{\bf q}(\mathbf Q,\omega),
\end{align}
where $\mathcal{T}_{0}=4\pi a/m$ is the scattering matrix of the bare $|c\rangle$-state excitations and $a$ is the corresponding scattering length. The pair propagator of two dark-state polaritons with a center of mass momentum ${\bf Q}$ and a relative momentum ${\bf q}$ is given by 
\begin{align}\label{PairProg}
\Pi_{\bf q}(\mathbf Q,\omega)=\left[\int\! \frac{{\rm }d\omega'}{2\pi}A_{c}(\mathbf q,\omega') G_{c}(\mathbf Q-\mathbf q,\omega-\omega')\right]+\frac{1}{2\epsilon_{\bf q}},
\end{align} 
and accounts for the dressing and broadening of $|c\rangle$-state by the light coupling as seen in Fig.~\ref{Fig2}. From the numerical solution of Eqs.~\eqref{Tm}-\eqref{PairProg},  we can obtain the total cross-section 
for the polariton collision $\sigma= m^2|{\mathcal T}|^2/2\pi$~\cite{PethickSmith2008book}. In the perturbative limit, the solution Eq.~(\ref{Tm}) recovers the known result,  
$\sigma=\sin^8\!\theta\:  8\pi a^2$, indicating that the polariton interaction is simply given by the interaction between the matter excitations weighted by their respective Hopfield coefficients $\sin^2\theta$~\cite{Fleischhauer2002}. 

\begin{figure}[b!]
\includegraphics[width=0.99\columnwidth]{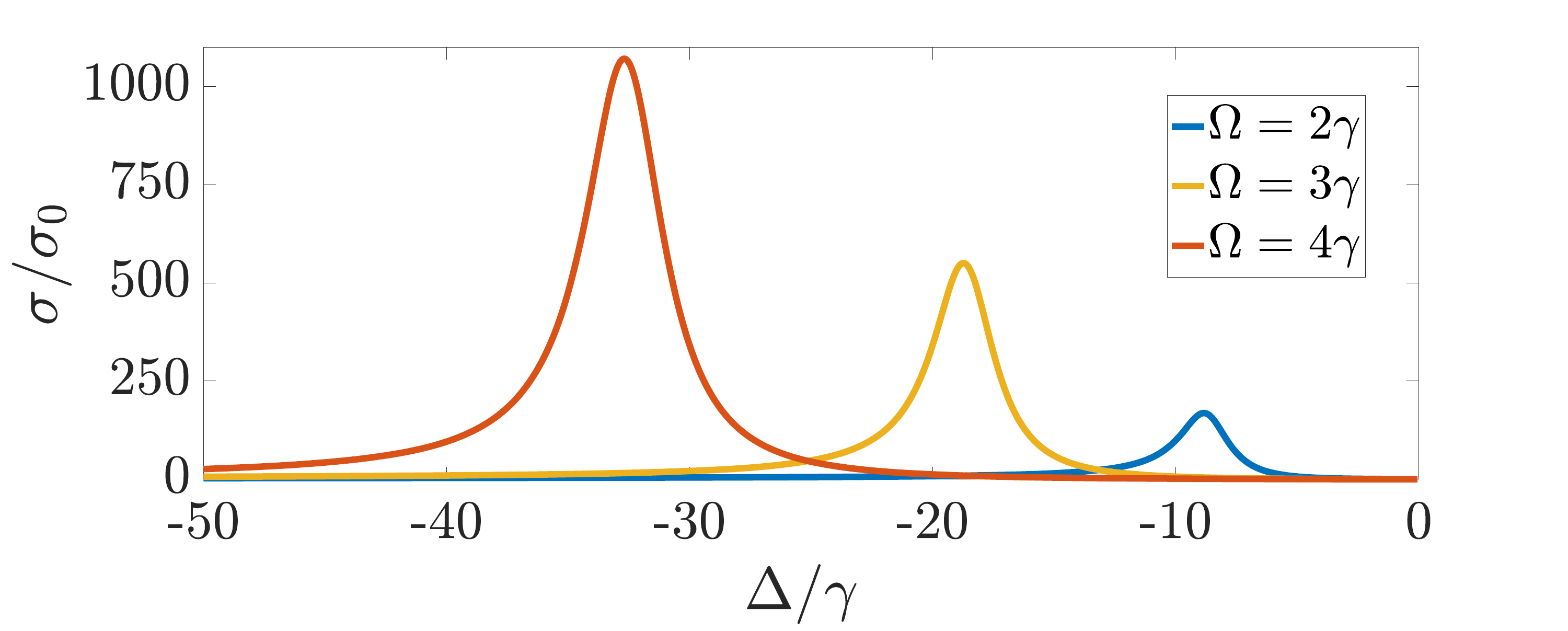}
\caption{Cross-section $\sigma/\sigma_0$ as a function of the single-photon detuning $\Delta$ for $\varepsilon_d=\gamma$, $\gamma_c=0.015\gamma$ and  different indicated values of the Rabi frequencies $\Omega$.}
\label{Fig3}
\end{figure}

The exact polariton scattering cross-section can be obtained from  Eqs.(\ref{Gc})-(\ref{PairProg}), and is shown in Fig.~\ref{Fig1}(c) as a function of $\Omega$. The center-of-mass momentum ${\bf p}_1+{\bf p}_2-2{\bf k}_{cl}$ of the $|c\rangle$ components is taken to be zero for simplicity. 
 We find a pronounced resonance around a given control field amplitude, where the polariton scattering cross-section is greatly enhanced compared to the collisional cross-section $\sigma_0=8\pi a^2$ of the bare excitations. 
This resonant behavior can be understood by noting that a positive scattering length $a$ implies a bound state of two bare $|c\rangle$-state excitations with a binding energy of $\varepsilon_d=1/ma^2$~\cite{Chin2010}.  
 The typical momentum states that make up this bound state lie well outside the EIT window. Their energy dispersion is, therefore, given by the high-momentum limit $p^2/2m-\Omega^2/\Delta$ shown in Fig.~\ref{Fig2}. 
 As the dimer consists of two such dressed $|c\rangle$ states, their energy  
\begin{align}
 E_D=2\epsilon_{c}+2\omega_{c}-\frac{1}{ma^2}-\frac{2\Omega^2}{\Delta},
\end{align}
is shifted by $-2\Omega^2/\Delta$.
On the other hand,  the energy of the two colliding dark-state polaritons is $2cp_r=2\epsilon_{c}+2\omega_{c}$. It follows that the 
collision is resonant when  $2cp_r=E_D$, yielding the condition  
 \begin{align}
 \frac{2\Omega^2}{\Delta}=-\frac{1}{ma^2}.
 \label{ResCon}
 \end{align}
 This simple resonance condition is shown in Fig.~\ref{Fig1}(c) and agrees rather well with the position of the resonances from the numerical calculations.
 
 While the position of the resonance can be well estimated from the simple relation~(\ref{ResCon}), the actual extent of the interaction enhancement  is determined by additional factors. As shown in Fig.~\ref{Fig3}, the enhancement of the polariton scattering cross section indeed becomes stronger with an increasing control-field Rabi frequency $\Omega$. This can be understood from the effect of the $|e\rangle$-state decay, which tends to broaden the resonance and thereby limits
 the achievable enhancement of $\sigma$. As we increase $\Omega$, it follows from Eq.~\eqref{ResCon} that the resonance occurs at 
 larger single-photon detuning $\Delta$. As a result, the $|e\rangle$-state is driven further away from resonance, which suppresses decay and leads to a larger enhancement of the scattering cross-section as observed in Fig.\ref{Fig3}.

\begin{figure}[t!]
\includegraphics[width=0.99\columnwidth]{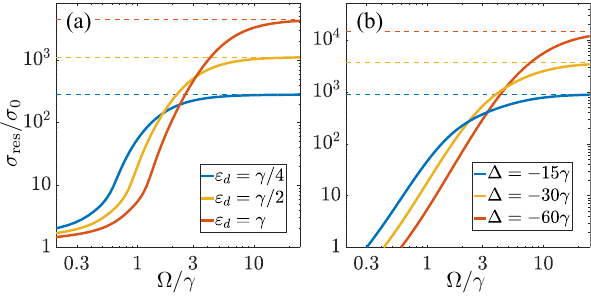}
\caption{(a) Resonant enhancement of the scattering cross-section,  $\sigma_{\rm res}/\sigma_0$, as a function of $\Omega$ for $\gamma_c=0.015\gamma$ and different indicated values of $\varepsilon_d$. Panel (b) shows $\sigma_{\rm res}/\sigma_0$ for $\gamma_c=0.015\gamma$ and different indicated values of $\Delta$. The horizontal dashed lines show the saturation limit, $\sigma_{\rm res}^{(\infty)}$, predicted by Eq.~\ref{sigma_max}.
}
\label{Fig4}
\end{figure}

We can analyse this effect more quantitatively by considering the resonant cross-section $\sigma_{\rm res}$, i.e. the maxima of the interaction resonances in Fig.~\ref{Fig3}. In Fig.~\ref{Fig4} we show $\sigma_{\rm res}$ as a function of $\Omega.$ Starting from small values of $\Omega$, the resonant cross-section initially increases with the control-field Rabi frequency but eventually saturates to an asymptotic value that depends on the dimer energy $\varepsilon_d$ and the decay rate $\gamma_c$ of the $|c\rangle$-state excitation. 

This behavior can be understood from the following estimate of the cross-section. First note that the main contribution from the pair propagator in Eq.(\ref{PairProg}) stems from high momenta, such that it is well approximated by
\begin{equation}
  \int\!\!\frac{{\rm d}^3 q}{(2\pi)^3}\Pi_{\bf q}\approx\frac{-im^{3/{2}}}{4\pi}\sqrt{\omega-\frac{Q^2}{4m}-2cp_r+2\frac{\Omega^2}{\Delta}+2i\tilde\gamma}.
  \label{PairPropVac}
  \end{equation}
  Here,  $\tilde\gamma=\gamma_c+(\Omega/\Delta)^2\gamma$ denotes the width of the dressed $|c\rangle$-state excitation, due to its direct decay and the small admixture, $\Omega^2/\Delta^2\ll1$, of the  $|e\rangle$-state excitation.
  Substituting the energy $\omega=2cp_r$ of the two colliding dark-state polaritons, $Q=0$, and the resonance condition Eq.~\eqref{ResCon} into 
  Eq.~\eqref{PairPropVac},  we obtain from  Eq.~\eqref{Tm} 
  $\mathcal T_0/\mathcal T_{cc}=1-\sqrt{1-2i\tilde\gamma/\varepsilon_d}$. 
 Expanding this  for $\tilde\gamma\ll \varepsilon_d$ yields
\begin{equation}
\sigma_{\rm res}=\left(\frac{\varepsilon_d}{\tilde{\gamma}}\right)^2\sigma_0=\left(\frac{\varepsilon_d}{\gamma_c}\frac{\Omega^2}{\Omega^2+\frac{\gamma}{4\gamma_c}\varepsilon_d^2}\right)^2\sigma_0.
\label{approx}
\end{equation}
Therefore, the enhancement increases rapidly as $\sigma_{\rm res}/\sigma_0\sim 4\Omega^4/(\gamma^2\varepsilon_d^2)$ with the control-field Rabi frequency and eventually saturates to a value of $\sim (\varepsilon_d/\gamma_c)^2$. 
For large control field Rabi frequency $\Omega$, the
polariton scattering cross-section thus saturates at the maximum value 
\begin{equation}
\sigma_{\rm res}^{(\infty)}= \frac{8\pi}{\gamma_c^2m^2a^2},
\label{sigma_max}
\end{equation}
that is limited only by the decay of the long-lived $|c\rangle$-state excitation. Remarkably, the maximum cross-section $\sigma_{\rm res}^{(\infty)}$ increases with decreasing scattering length $a$, thereby generating strongly interacting polaritons in a  weakly interacting optical medium. The origin of this intriguing effect can be understood from expanding the scattering matrix around the Feshbach pole yielding  
${\mathcal T}(0,2cp_r+\omega)\approx \frac{8\pi}{m^2a}\frac{1}{\omega+2i\tilde\gamma}$. Thus, the residue of the  pole is proportional to  $1/a$~\cite{Bruun2010,SM}. Physically, this scaling breaks down when the dimer state probes length 
scales of the order of the range $r_0$ of the interaction, i.e. for $a\lesssim r_0$ where  the momentum dependence of the interaction becomes important. Equation \eqref{approx} remains valid for general interactions with a non-zero range, since it follows from a pole expansion of the $\mathcal T$-matrix around the bound state energy.

Cold atomic gases offer an ideal optical medium \cite{hau1999,Bajcsy2009,Riedl2012,Hsiao2018} to explore this surprising behavior. Consider a cold gas of $^{87}$Rb atoms as a typical example. While the scattering length of rubidium atoms can be tuned via magnetic Feshbach resonances \cite{verhaar2002}, already the field-free value corresponds to a rather large dimer energy of $\varepsilon_d/2\pi\sim 4$MHz. With a realistic linewidth $\gamma_c/2\pi\sim100$kHz of the two-photon transition, Eq.(\ref{approx}) predicts a greatly enhanced cross-section $\sigma/\sigma_0\sim1600$. Such a drastic enhancement will significantly increase the nonlinear response of the gas, and should thus be observable in systematic measurements of the nonlinear refraction and absorption of slow light under EIT conditions. 


Enhancing the effective interactions between photons is broadly important in solid-state optics. Hereby, recently explored two-dimensional van der Waals heterostructures \cite{Liu2020} offer a flexible approach to implementing the required three-level driving scheme. 
Using moir\'e lattices in twisted bilayer transition metal dichalcogenides, cavity photons can be coupled directly to intralayer excitons with a Rabi frequency of $15-50\text{meV}$~\cite{Zhang2021}. These intralayer excitons represent the $|e\rangle$-state excitations introduced above. They can hybridize via a coherent electron tunneling with interlayer excitons ~\cite{Togan2018,Hsu2019,Brem2020}, which 
can be tuned by changing the relative angle between the two layers~\cite{Alexeev2019}. 
This coherent tunneling will likely  be dominant compared to incoherent tunneling involving phonons, when the valence bands of the two layers are close in energy~\cite{Jiang2021,Alexeev2019}.
 Interlayer excitons  implement the meta-stable  $|c\rangle$-state excitations, since they are remarkably long lived with a decay rate $\gamma_c\sim 1\mu$eV \cite{rivera2015} that is much smaller than that of the intralayer excitons. Moreover, the band structure generated 
 by the moir\'e lattice combined with  
 the  sizable dipole-dipole interaction inherited from the interlayer excitons leads to the emergence of a repulsive bound state~\cite{Winkler2006wu}.

Cuprous oxide is another well-studied material in which a three-level coupling scheme could be implement. Hereby, the incident probe photons can be tuned to generate 2$p$-excitons of the yellow series, while a strong additional field can be used to drive the control-transition to long-lived 1s-para excitons \cite{Brandt2007} in order to establish EIT \cite{Artoni2000}. The yellow Rydberg series of $p$-state excitons in cuprous oxide has been studied in a number of recent experiments \cite{Kazimierczuk2014,Steinhauer2020,heckotter2021,Orfanakis2021,Versteegh2021}. Since this transition is dipole-forbidden, the narrow excitation lines come at the cost of a relatively low light-matter coupling strengths. Optical microcavities can, however, be used to enhance the photon-coupling and generate exciton polaritons in Cu$_2$O, while the implementation of EIT can suppress phonon-induced absorption and decoherence \cite{Walther2020}. From the collisional cross-section of $\sigma_0\sim50$nm$^2$ for $1s$-para excitons \cite{Shumway2001,Yoshioka2011} we estimate $\varepsilon_d\sim 20$meV, which would yield a large polariton-scattering length of several $\mu$m for $1s$-exciton linewidths, $\gamma_c$, in the $\mu$eV range. Indeed, linewidths as low as $\gamma_c\sim5$neV have been observed in experiments \cite{Koirala2013}. Such large polariton scattering lengths, well above the wavelength of the probe photons, would reach the strong-blockade regime for confined polaritons \cite{Delteil2019} and thereby generate optical nonlinearities at the level of individual photons.

Note that as opposed to earlier schemes to realise strong photon-photon interactions using Feshbach resonances between polaritons~\cite{Takemura2014,Wouters2007}, it is the molecule level and not the polaritons that is tuned into resonance in our three-level scheme. Thus, the composition of the polaritons is fixed, which ensures lossless propagation and an efficient photon-photon scattering. In addition, our scheme enables coupling to long-lived Feshbach dimers of indirect excitons in contrast to earlier experiments, where the resonance was strongly reduced due to fast decay of short lived  Feshbach bi-exciton dimers~\cite{Takemura2014}. 

In conclusion, we have presented an exact study of polariton scattering in an optical three-level medium with short-range interactions. 
Our results show that the resulting effective interaction between dark-state polaritons can be orders of magnitude larger than the underlying interaction between their matter constituents. This effect emerges from the two-photon coupling and bound dimer states of long-lived matter excitations and applies to a range of systems from atomic ensembles to excitons in semiconducting materials. While we have focused here on the free-space propagation of light in three dimen- sions, such applications motivate future investigations into specific implementations, including two-dimensional geometries of optical cavities, band structure effects in moir\'e superlattices, finite-range exciton interactions and decoherence in solid-state systems. The possibility to control and enhance polariton interactions in otherwise weakly interacting systems holds exciting perspectives for optical science, from few-photon applications in optical quantum technologies \cite{Chang2014} to fundamental explorations of strongly interacting photonic many-body systems \cite{Chang2008,Konishi2021,Roux2020}. 

We acknowledge financial support from the Danish National Research Foundation through the Center of Excellence “CCQ” (Grant agreement no.: DNRF156), the Carlsberg Foundation through the 'Semper Ardens' project QCooL, 
the Villum Foundation, and the Independent Research Fund Denmark - Natural Sciences via Grant No.\ DFF - 8021- 00233B.  A.C.G acknowledges the Theory of Condensed Matter Department, University of Cambridge for financial support.  ACG acknowledges grant No. IN108620 from DGAPA (UNAM)

\bibliography{Polariton}

\begin{thebibliography}{59}
\bibitem [{Rb8()}]{Rb87}%
  \BibitemOpen
  \href@noop {} {}\bibinfo {note} {Specifically, we consider the D$_2$ line of
  Rb$^{87}$, for which $\lambda_c=2\pi/k_c=780$nm and $\gamma/2\pi=6.1$MHz,
  such that $\gamma/(mc^2)\approx3.1\times10^{-19}$ and $c
  k_c/\gamma=6.3\times10^{7}$.}\BibitemShut {Stop}%
\end{thebibliography}
\end{document}